\documentclass[12pt,nohyper,notoc]{JHEP}

\usepackage{amsmath,amssymb,cite,graphicx} 
\usepackage{epsf}
	
\newcommand{\beq}{\begin{eqnarray}}
\newcommand{\eeq}{\end{eqnarray}}

\newcommand{\centeron}[2]{{\setbox0=\hbox{#1}\setbox1=\hbox{#2}\ifdim
                           \wd1>\wd0\kern.5\wd1\kern-.5\wd0\fi \copy0
                           \kern-.5\wd0\kern-.5\wd1\copy1\ifdim\wd0>\wd1
                           \kern.5\wd0\kern-.5\wd1\fi}}
\newcommand{\ltap}{\>\centeron{\raise.35ex\hbox{$<$}}
                   {\lower.65ex\hbox{$\sim$}}\>}
\newcommand{\gtap}{\>\centeron{\raise.35ex\hbox{$>$}}
                   {\lower.65ex\hbox{$\sim$}}\>}

\newcommand{\lsim}{\mathrel{\ltap}}

\title{Constraints on the SU(3) Electroweak Model}

\author{Csaba Cs\'aki$^{a}$, 
Joshua Erlich$^b$, Graham D. Kribs$^{c}$, and John Terning$^b$\\
$^a$Newman Laboratory of Nuclear Studies, Cornell University, Ithaca, NY 14853,
USA
\\
$^b$Theory Division T-8, Los Alamos National Laboratory, Los Alamos,
NM 87545, USA
$^c$Department of Physics, University of Wisconsin, Madison, WI 53706, 
USA \vspace*{0.2in} \\
\centerline{\phantom{0}}
\centerline{\tt csaki@mail.lns.cornell.edu, erlich@lanl.gov, 
kribs@pheno.physics.wisc.edu,} \centerline{\tt terning@lanl.gov}}

\abstract{
We consider a recent proposal by Dimopoulos and Kaplan to embed the
electroweak SU(2)$_L$ $\times$ U(1)$_Y$ into a larger group SU(3)$_W$
$\times$ SU(2) $\times$ U(1) at a scale above a TeV\@.  This idea is
motivated by the prediction for the weak mixing angle $\sin^2 \theta_W
= 1/4$, which naturally appears in these models so long as the
gauge couplings of the high energy SU(2) and U(1) groups are
moderately large.  The extended gauge dynamics results in new
effective operators that contribute to four-fermion interactions and
$Z$ pole observables.  We calculate the corrections to these
electroweak precision observables and carry out a global fit of the
new physics to the data.  For SU(2) and U(1) gauge couplings larger
than $1$, we find that the $95$\% C.L. lower bound on the matching (heavy
gauge boson mass) scale is $11$ TeV\@.  We comment on the fine-tuning
of the high energy gauge couplings needed to allow matching scales
above our bounds.  The remnants of SU(3)$_W$ breaking include
multi-TeV SU(2)$_L$ doublets with electric charge ($\pm 2,\pm 1$).
The lightest charged gauge boson is stable, leading to cosmological
difficulties.}

\preprint{{\tt hep-ph/0204109}}

\begin{document}

\section{Introduction}

Unification of the standard model forces into a larger gauge
symmetry is perhaps the most elegant physics beyond the Standard Model.
Full unification of SU(3)$_c$ $\times$ SU(2)$_L$ $\times$ U(1)$_Y$
into a grand unified theory such as SU(5) simplifies the matter content of 
the model, determines the hypercharge normalization, and predicts 
gauge coupling unification.  For the standard model, the combination of 
precision measurements of the low energy gauge couplings plus the 
non-observation of proton decay rule out all of the simplest 
non-supersymmetric grand unified theories.  Given this disappointing result, 
it is natural to 
consider other alternatives, such as the partial unification of 
SU(2)$_L$ $\times$ U(1)$_Y$ into a larger group.  Weinberg first
considered unifying the electroweak gauge symmetries into SU(3)
back in 1972 \cite{Weinberg}.  The left-handed doublet 
plus the right-handed singlet leptons fit neatly into triplets of 
SU(3) \cite{KonopinskiMahmoud}.  The U(1)$_Y$ normalization 
is fixed by the embedding 
into SU(3), which results in the famous prediction $\sin^2 \theta_W = 1/4$
at the scale of SU(3) breaking.  However, quarks cannot be simply 
embedded into SU(3) representations due to their fractional 
hypercharge.

Recently there has been renewed interest in predicting
$\sin^2 \theta_W = 1/4$ through new gauge dynamics that appears not 
too far above the electroweak scale \cite{DK1,others}.  
The proposal by Dimopoulos and Kaplan \cite{DK1} embeds 
the electroweak SU(2)$_L$ $\times$ U(1)$_Y$ into a larger product gauge 
symmetry SU(3)$_W$ $\times$ SU(2) $\times$ U(1) in which quarks and 
leptons are charged under just the high energy SU(2) $\times$ U(1).
The larger gauge structure is spontaneously broken down to the 
electroweak model by the vev of a scalar field $\Sigma$ 
transforming under all of the gauge groups.  By choosing a larger 
product gauge group that includes SU(2) $\times$ U(1), there is
no difficulty in accommodating all quark and lepton hypercharges
(since the quark and lepton hypercharges simply 
corresponds to their  charges under the 
high energy U(1)).  The price to be paid for this freedom is
twofold:  The normalization of hypercharge is unexplained, and
there is no simplification of the matter content of 
the standard model. 
However, if the gauge couplings of the high energy SU(2) and U(1) groups 
are at least moderately large, the electroweak gauge couplings are
determined primarily by SU(3)$_W$ with the Weinberg $\sin^2 \theta_W = 1/4$
prediction of the weak-to-hypercharge coupling ratio.

Since the minimal gauge extension to SU(3)$_W$ $\times$ SU(2) $\times$ 
U(1) has nothing to do with the U(1)$_Y$ normalization, the intersection 
of the SU(2)$_L$ and U(1)$_Y$ gauge couplings
to any value at any scale is possible.  Only when the high energy U(1) 
is embedded into a simple group at an even higher scale is the 
normalization fixed.  The simplest possibility is that the U(1)
embedded into an SU(2) group \cite{DK1}.  This determines the
U(1) normalization and fixes the $\Sigma$ field's U(1) charge to
be $1/2$.  From this point on, we tacitly assume this assignment of
U(1) charge to the $\Sigma$ field.

The breaking of SU(3)$_W$ $\times$ SU(2) $\times$ U(1) to the
electroweak model results in eight new gauge bosons that obtain a mass 
of order $\langle \Sigma \rangle$ times gauge couplings, and four 
that become the
electroweak gauge bosons.  Integrating out the heavy gauge bosons
results in new effective operators suppressed by squares of gauge boson 
masses that contribute to four-fermion processes and $Z$ pole
observables \cite{STC,Burgess}.
In this paper we calculate the mass spectrum and the low-energy 
effective Lagrangian of the SU(3)$_W$ $\times$ SU(2) $\times$ U(1) 
model and the resulting corrections to electroweak precision 
observables due to the the heavy gauge bosons.

We may gain some intuition as to what we might expect
by remembering an older idea in which the left-handed quarks and left-handed 
leptons 
are charged under two different SU(2) gauge symmetries \cite{Georgiununified}.
In this ``ununified standard model'', the scale of the SU(2)$_q$ $\times$ 
SU(2)$_l$ breaking was found to be significantly 
constrained by electroweak precision corrections \cite{CST}.
The limit on the matching scale rises dramatically for the case 
in which the leptonic coupling is taken to be much larger 
than the quark coupling $g_l \gg g_q$.  This is because the matching 
scale for the new gauge dynamics corresponds to the heaviest 
gauge boson, of order $g_l u$ where $u$ is the vev of a bidoublet
scalar field that breaks SU(2)$_q$ $\times$ SU(2)$_l$ to SU(2)$_L$.

It is this general observation suitably applied to the SU(3)$_W$
$\times$ SU(2) $\times$ U(1) model that significantly constrains the
new gauge dynamics.  We emphasize that the matching scale is by
definition the scale at which the high energy theory including the
full SU(3)$_W$ gauge symmetry is decomposed into the standard model
SU(2)$_L$ $\times$ U(1)$_Y$.  Effective theory demands that this scale
coincide with the scale of the heaviest gauge boson, since the full
product gauge theory cannot be realized until all of the gauge bosons
are propagating degrees of freedom.  Broken gauge multiplets that are
split in mass due to a hierarchy in gauge couplings are accounted for
through threshold corrections to the renormalization group evolution
of the electroweak gauge couplings.

The organization of this paper is as follows.  In Sec.~\ref{spectrum-sec} 
we calculate the mass spectrum of the SU(3)$_W$ $\times$ SU(2) $\times$ U(1) 
broken to SU(2)$_L$ $\times$ U(1)$_Y$.  After integrating out the 
heavy gauge bosons, we calculate their tree-level effects on the masses
and couplings between the electroweak gauge bosons and matter in
Sec.~\ref{effective-sec}.  The resulting modifications to the electroweak
theory are used to calculate the corrections to precision electroweak 
observables in Sec.~\ref{observables-sec}, which are determined 
by just two parameters.  We then perform 
a global fit of the new physics contributions to the experimental 
observables.  The constraints on the new physics extracted from the fit 
imply constraints on the region of allowed SU(2) $\times$ U(1)
gauge couplings.  This requires evolving the electroweak gauge couplings 
using the renormalization group equations with thresholds corrections
that we calculate in Sec.~\ref{RG-sec}.
In Sec.~\ref{pheno-sec} we discuss the phenomenology 
of the heavy gauge bosons.  We focus on the electrically charged gauge
bosons that do not couple to quarks and leptons, pointing out that they are 
only produced in pairs and that the lightest one does not decay.  
We estimate the
cosmological abundance of these charged, stable relics and find 
that it is of order the critical density or larger.  In addition to
resulting cosmological difficulties, this is problematic
in light of the strong experimental bounds on the abundance of charged 
stable particles through searches for heavy isotopes of ordinary
nuclei.  Finally, we conclude in Sec.~\ref{conclusions-sec}.

\section{Spectrum}
\label{spectrum-sec}

The gauge group of the model is SU(3)$_W$ $\times$ SU(2) $\times$ U(1)
with gauge couplings $g_3$,  $\tilde g$, and $\tilde g^\prime$ respectively.
The quarks and leptons are uncharged under SU(3)$_W$, having the
same quantum numbers under SU(2) $\times$ U(1) as they do under
SU(2)$_L$ $\times$ U(1)$_Y$.
The Higgs field is replaced by two scalars $\Phi$ and $\Sigma$.
$\Phi$ is uncharged under SU(3)$_W$, but with the same quantum numbers 
as the Higgs under SU(2) $\times$ U(1), while $\Sigma$ 
transforms as $(3,2,-1/2)$.
At some high scale SU(3)$_W$ $\times$ SU(2) $\times$ U(1) is broken down to
the electroweak group SU(2)$_L$ $\times$ U(1)$_Y$ by the vev
\beq
\langle\Sigma\rangle=\left(\begin{array}{cc} M & 0 \\ 0 & M \\ 0 & 0
\end{array}\right).
\eeq
The gauge structure of this model is similar to the recently proposed 
deconstructed models where the Higgs is a pseudo-Goldstone boson~\cite{Nima}.

The gauge bosons of the three groups mix to form the following representations
of SU(2)$_L$ $\times$ U(1)$_Y$: $(3,0) \oplus (3,0) \oplus (1,0)\oplus(1,0)
\oplus (2,3/2) \oplus (2,-3/2)$. We first consider the $(3,0)$ sector. These
fields arise from the SU(2) gauge bosons $\tilde{W}^a$ and the 
$A^{1,2,3}$ gauge bosons of the SU(2) subgroup of SU(3)$_W$.
In the ($A^a$, $\tilde W^a$) basis (for $a=1,2,3$)
the mass matrix is:
\beq
M^2 \left(\begin{array}{cc} g_3^2  &  g_3 \tilde g  \\  
 g_3 \tilde g  &\tilde g^2   \end{array}\right)
\eeq
Thus the light and heavy mass eigenstates  are:
\beq
W_L^a &=& c_\phi A^a - s_\phi \tilde W^a\\
W_H^a &=& s_\phi A^a + c_\phi \tilde W^a
\eeq
with masses
\beq
M_{W_L} &=& 0 \\
M_{W_H} &=& \sqrt{\tilde g^2 + g_3^2} M 
\eeq
where
\beq
s_\phi= \frac{g_3}{\sqrt{\tilde g^2+g_3^{2}}},\ \ \ \ 
c_\phi=\frac{\tilde g}{\sqrt{\tilde g^2+g_3^{2}}} ~.
\eeq

The SU(2) singlets arise from the U(1) gauge boson $\tilde{B}$ and 
the $A^8$ component of the SU(3)$_W$ gauge bosons.  These will constitute 
the $(1,0)$ sector.  The mass matrix in the ($A^8$,$\tilde{B}$) basis 
at the high scale is
\beq
M^2 \left(\begin{array}{cc} \frac{1}{3} g_3^2 &
   \frac{1}{\sqrt{3}} g_3 \tilde g^\prime  \\  
  \frac{1}{\sqrt{3}} g_3 \tilde g^\prime  
&\tilde g^{\prime\,2} \end{array}\right)
\eeq
Thus the light and heavy mass eigenstates  are:
\beq
B_L &=& c_\psi A^8 - s_\psi \tilde B\\
B_H &=& s_\psi A^8 + c_\psi \tilde B
\eeq
with masses
\beq
M_{B_L} &=& 0 \\
M_{B_H} &=& \sqrt{\tilde g^{\prime\,2} + \frac{g_3^2}{3}} M
\eeq
where
\beq
s_\psi= \frac{g_3}{\sqrt{3 \tilde g^{\prime\,2}+g_3^{2}}},\ \ \ \ 
c_\psi=\frac{\sqrt{3} \tilde g^\prime}{\sqrt{3 \tilde g{\prime ^2}+g_3^{2}}} ~.
\eeq

Finally we consider the $(2,\pm 3/2)$ sector that comes from the
$A^{4,5,6,7}$ gauge bosons of SU(3)$_W$.  These fields have no 
SU(2) $\times$ U(1) partners to mix with, and thus their mass is simply
\beq
M_i = \frac{g_3}{\sqrt{2}} M ~.
\eeq
Since $g_3$ is expected to be smaller than $\tilde g$ and $\tilde g^\prime$,
these gauge bosons will be substantially
lighter than $W^a_H$ and $B_H$.  These intermediate scale gauge bosons, 
however,
do not have any direct couplings to quarks and leptons.  Thus they
would only be seen in processes involving virtual SM gauge bosons.

The effective gauge couplings of the SU(2)$_L$ $\times$ U(1)$_Y$ groups are:
\beq
g &=& \tilde g s_\phi\\
g^\prime &=& \tilde g^\prime s_\psi
~.
\eeq
The coupling of $W^a_H$ $(B_H)$ to quarks and leptons is 
$- \tilde g c_\phi$ $(- \tilde g^\prime  c_\psi)$.

\section{The Low-energy Effective Action}
\label{effective-sec}

We now construct the effective theory below the mass scale of the heavy
gauge bosons.  Integrating out $W^a_H$ and $B_H$ induces  additional 
operators in the effective theory. These operators modify the usual
relations between the standard model parameters, and therefore their
coefficients can be constrained from electroweak precision measurements.
There are three types of operators that will be relevant for us:
corrections of the coupling of the SU(2)$_L$ $\times$ U(1)$_Y$ gauge
bosons to their corresponding currents, operators that are quadratic 
in the SU(2)$_L$ $\times$ U(1)$_Y$ gauge fields and quartic in the ordinary
Higgs field $\Phi$, and four-fermion operators.

Exchanges of the  heavy $W^a_H$ and $B_H$ gauge bosons give the 
following operators which are quadratic in the light gauge fields:
\begin{eqnarray}
{\cal L}_{2}&=&
-\frac{g^2 c_\phi^4 }{8 M^2} W^{a \mu}_L W^{a}_{\mu L}
(\Phi^\dagger \Phi )^2  - \frac{g^{\prime\,2}  c_\psi^4}
{8 M^2} B^{\mu}_L B_{\mu L} (\Phi^\dagger \Phi )^2
-\frac{g^2 c_\psi^4 }{8 M^2} W^{a \mu}_L W^{b}_{\mu L}
(\Phi^\dagger \sigma^a  \Phi ) (\Phi^\dagger  \sigma^b \Phi ) \nonumber \\
&&
- \frac{g^{\prime\,2}  c_\phi^4}
{8 M^2} B^{\mu}_L B_{\mu L}
(\Phi^\dagger  \sigma^a \Phi )^2 
- \frac{g g^{\prime } 
 (c_\phi^4+c_\psi^4)}{4 M^2} W^{a \mu}_L B_{\mu L}
(\Phi^\dagger \sigma^a  \Phi)(\Phi^\dagger \Phi ), 
\label{operators}
\end{eqnarray}
where $\sigma^a$ are the Pauli $\sigma$ matrices. 
For example, the first term arises in the following way. The kinetic term of
the Higgs field $(D_\mu \Phi)^\dagger D^\mu \Phi$ contains the coupling
\beq
{\cal L}_{\tilde W^2 \Phi^2}=
\frac{\tilde{g}^2}{4}  \tilde{W}^a_\mu \tilde{W}^{b \mu} (\Phi^\dagger 
\sigma^a \sigma^b  \Phi )~.
\eeq
Expressing
$\tilde{W}^a = c_\phi W_H^a-s_\phi W_L^a$ we obtain a coupling 
between the heavy and light gauge bosons of the form 
\beq
{\cal L}_{W_L W_H \Phi^2}=-\frac{\tilde{g}^2 s_\phi 
c_\phi}{4}  (W^a_{\mu L}W^{b \mu}_H+W^b_{\mu L}W^{a \mu}_H)
(\Phi^\dagger 
\sigma^a \sigma^b \Phi )= -\frac{\tilde{g}^2 s_\phi 
c_\phi}{2}  W^a_{\mu L}W^{a \mu}_H
(\Phi^\dagger 
\Phi )~.
\eeq 
The first term in ${\cal L}_{2}$ then arises by integrating out the
heavy gauge boson $W^{a \mu}_H$ by taking its equation of
motion and expressing it in terms of the light fields.
The operators in (\ref{operators}) 
are the ones that give corrections to the light
gauge boson masses after $\Phi$ gets a vev. Thus
after $\Phi$ gets the usual vev:
\beq
\langle \Phi \rangle = 
\left(\begin{array}{c} 0 \\ v/ \sqrt{2}\end{array}\right)~,
\eeq 
and including the effects of the higher dimension operators 
(\ref{operators}), we find
that the mass of the $W$ is
\beq
M_W^2 &=&  g^2  \frac{v^2}{4}\left(1
-\frac{  c_\phi^4 v^2}{ 4  M^2}\right)
\eeq
The mass matrix in the ($W^3_L$,$B_L$) basis
 is:
\beq
\frac{v^2}{4}\left(1
-\frac{  (c_\psi^4+ c_\phi^4) v^2}{ 4  M^2}\right)
\left(\begin{array}{cc} g^2  & - g g^\prime
  \\  
- g g^\prime
  & g^{\prime\,2}  \end{array}\right)
\eeq
So the mass of the $Z$ is
\beq
M_Z^2 &=&(g^2 + g^{\prime\,2})\frac{v^2}{4} 
\left(1-\frac{ ( c_\phi^4 +c_\psi^4)v^2}{4 M^2}
 \right)
\eeq

In addition, exchanges of $W^a_H$ and $B_H$ give
corrections to the coupling of the SU(2)$_L$ $\times$ U(1)$_Y$ gauge
bosons to their corresponding currents and additional four-fermion operators:
\beq
{\cal L}_{\rm c}&=&
g W^a_{L \mu} J^{a \mu}\left(1- (\Phi^\dagger \Phi )
  \frac{c_\phi^4}{2 M^2}\right)
+  g^\prime B_{L \mu} J^\mu_{Y}
\left(1-  (\Phi^\dagger \Phi ) \frac{c_\psi^4}{2 M^2}\right)
\nonumber \\
&&-  g  W^a_{L \mu}J^\mu_{Y} (\Phi^\dagger \sigma^a \Phi ) \frac{c_\psi^4}{2 M^2}
-  g^\prime B_{L \mu} J^{a \mu} (\Phi^\dagger \sigma^a \Phi )
 \frac{c_\phi^4}{2 M^2}-J_\mu^aJ^{a\mu} \frac{c_\phi^4}{2 M^2}-J_\mu^Y
  J^{Y\mu}\frac{c_\psi^4}{2 M^2}
\label{currents}
\eeq

Using this expression we can now evaluate the effective Fermi coupling
$G_F$ in this theory. The simplest way to obtain the answer for this is by
integrating out the $W_L$ bosons from the theory by adding the W mass term
to (\ref{currents}). The expression we obtain for the effective four-fermion 
operator is
\begin{equation}
-\frac{g^2}{2M_W^2} J^{+\mu}J^-_\mu (1-\frac{c_\phi^4 v^2}{2M^2})-
J^{+\mu}J^-_\mu \frac{c_\phi^4}{2M^4}=-2 \sqrt{2} G_F J^{+\mu}J^-_\mu,
\end{equation}
where $J^\pm=\frac{1}{2}(J^1\pm iJ^2)$.
Plugging in the correction to the W mass we obtain that $G_F$ in this model
is uncorrected, that is 
\begin{equation}
\label{treeGF}
G_F= \frac{1}{\sqrt{2}v^2}.
\end{equation}
This is in fact not a 
coincidence, but a general result in such models which was first 
derived in 
\cite{Georgiununified}.  The ($A^a$, $\tilde{W}^a$)
mass matrix can be written as a product of the coupling matrix $G$ 
and the matrix of vevs $V$: \beq
{\cal M}^2=G\,V^2\,G, \eeq
where $G={\rm diag} (g_3,\tilde{g})$ 
and the matrix of squared vevs is, \beq
V^2=M^2\left( \begin{array}{cc}
1 & 1 \\
1 & (1 + \frac{v^2}{4M^2}) \end{array}
\right)\, . \eeq
Then the charged current interactions at zero momentum transfer are given
by, \beq
\frac{1}{2}J_\mu^\dagger\,G{\cal M}^{-2}G\,J^\mu =\frac{1}{2}J_\mu^\dagger
\,V^{-2}\, J^\mu, \eeq
where $J^\mu=(0,j_q^\mu+j_l^\mu)$
is the charged quark $(j_q^\mu)$ and lepton $(j_l^\mu)$ 
current vector in the (SU(3)$_W$, SU(2)) basis.
Evaluating, \beq
V^{-2}=\frac{4}{v^2}\left(\begin{array}{cr}
(1+\frac{v^2}{4M^2}) & -1 \\
-1 & 1 \end{array}\right). \eeq
We find that the charged current four-fermion interaction is given by, \beq
\frac{2}{v^2}(j_q^\mu+j_l^\mu)^2.\eeq
From this we read off that $G_F$ is given by (\ref{treeGF}) and does not 
receive tree level corrections from the SU(3)$_W$ interactions.

Finally, to fix all SM parameters we need to identify the 
photon and the neutral-current couplings from (\ref{currents}):
\beq
{\cal L}_{\rm nc}=&&
e A_{\mu} J^\mu_{Q}
+ \frac{e}{s c}Z_\mu\left[ J^{3 \mu}
\left(1-  \frac{(c_\phi^4+c_\psi^4)v^2}{4 M^2}\right)
- J^{\mu}_Q\left(s^2 -  \frac{c_\psi^4 v^2}{4 M^2}\right)\right]\nonumber \\
&&-J^{3\mu}J^3_\mu \frac{c_\phi^4}{2M^2}-(J^3-J_Q)^\mu(J^3-J_Q)_\mu 
\frac{c_\psi^4}{2M^2}.
\label{nc}
\eeq
Here $e= gg'/\sqrt{g^2+g^{\prime\,2}}$ as in the standard model, 
thus there is no
correction to the expression of the electric charge $e$ 
compared to the SM\@. Similarly to the evaluation of the 
effective $G_F$, we can calculate the low-energy effective four-fermion 
interactions from the neutral currents. The result we obtain is
\begin{eqnarray}
{\cal L}_{NC}^{(4f)} = -\frac{2}{v^2} (J^3-s^2 J_Q)^2 -\frac{1}{2 M^2} (J_Q)^2 
\Big(s^4(c_\phi^4+c_\psi^4) -2 s^2 c_\psi^4 +c_\psi^4 \Big)~,
\end{eqnarray}
where the first term is just the SM result, while the second term is the 
correction. Note that the correction term to the four-fermion interactions 
contains only the charged currents, and so it does not contribute to 
neutrino scattering processes or atomic parity violation.

\section{The Contributions to Electroweak Observables}
\label{observables-sec}

To relate our parameters to observables we use the 
standard definition of
$\sin \theta_0$ from the $Z$ pole \cite{Sformulas},
\beq
\label{s2}
\sin^2 \theta_0 \cos^2 \theta_0 &=& \frac{\pi \alpha(M_Z^2)}{\sqrt{2} G_F M_Z^2}~,\\
\sin^2 \theta_0&=&0.23105 \pm 0.00008
\eeq
where $\alpha(M_Z^2)^{-1}=128.92\pm 0.03$ is the running SM fine-structure
constant evaluated at $M_Z$~\cite{ErlerLang}.
We can relate this measured value with the bare value 
in this class of models, 
\beq
\sin^2  \theta_0 = s^2 +\frac{s^2 c^2}{c^2-s^2} 
\frac{ ( c_\phi^4 +c_\psi^4)v^2}{4 M^2},
\label{rens2}
\eeq
which is obtained by considering all corrections to (\ref{s2}) in the 
usual way (see~\cite{Sformulas}).
Also, we have the simple result that the
running couplings defined by Kennedy and Lynn~\cite{Lynn}
which appear in $Z$-pole 
asymmetries are the same as
the bare couplings:
\beq
s^2_*(q^2)=s^2, \ \ \ 
e^2_*(q^2)=e^2 ~.
\eeq

In order to compare to experiments, we can relate our corrections of
the neutral-current couplings to the generalized modifications of the 
$Z$ couplings as defined by Burgess et al. \cite{Burgess},
\begin{equation}
\Delta {\cal L}=\frac{e}{sc} \sum_i \bar{f}_i \gamma^\mu 
(\delta \tilde{g}_L^{ff} \gamma_L + \delta \tilde{g}_R^{ff} \gamma_R) f_i 
Z_\mu,  \end{equation}
where 
\begin{equation}
\frac{1}{s c} = \frac{1}{s_0 c_0} 
\left[ 1 + \frac{(c_\phi^4 + c_\psi^4)v^2}{8 M^2} \right] \; .
\end{equation}
 From (\ref{nc}) we obtain that\footnote{Note that the correction to
the $Z$ coupling $\tilde{g}^{ff}$ should not be confused with the
high energy SU(2) $\times$ U(1) gauge couplings $\tilde{g},\tilde{g}'$.}
\begin{equation}
\delta \tilde{g}^{ff}=\frac{v^2}{4M^2} 
\left[q^f c_\psi^4-t_3^f (c_\phi^4+c_\psi^4)\right].
\end{equation}
For the individual couplings this implies
\begin{eqnarray}
\delta \tilde{g}_L^{uu}&=& 
  \frac{-3c_\phi^4+c_\psi^4}{24 M^2}v^2=\frac{1}{6}(-3c_1+c_2),
\qquad\>\> \delta \tilde{g}_R^{uu}= 
  \frac{c_\psi^4}{6 M^2}v^2=\frac{2}{3}c_2, \nonumber \\
\delta \tilde{g}_L^{dd}&=&
  \frac{3c_\phi^4+c_\psi^4}{24 M^2}v^2=\frac{1}{6} (3 c_1+c_2),
\qquad\qquad \delta \tilde{g}_R^{dd}= 
 -\frac{c_\psi^4}{12 M^2}v^2=-\frac{1}{3} c_2, \nonumber \\
\delta \tilde{g}_L^{ee}&=& 
  \frac{c_\phi^4-c_\psi^4}{8M^2}v^2 =\frac{1}{2}(c_1-c_2), 
\qquad\qquad\quad \delta \tilde{g}_R^{ee}= 
  -\frac{c_\psi^4}{4M^2}v^2=-c_2, \nonumber \\
\delta \tilde{g}_L^{\nu\nu}&=&
  -\frac{c_\phi^4+c_\psi^4}{8M^2}v^2 = -\frac{1}{2}(c_1+c_2),
\end{eqnarray}
where $\delta \tilde{g}^{\mu\mu}=\delta \tilde{g}^{\tau\tau}=
\delta \tilde{g}^{ee}$, and similarly 
$\delta \tilde{g}^{tt}=\delta \tilde{g}^{cc}=\delta \tilde{g}^{uu}$, 
$\delta \tilde{g}^{bb}=\delta \tilde{g}^{ss}=\delta \tilde{g}^{dd}$. 
We have introduced the notation
\begin{equation}
c_1 = \frac{c_\phi^4 v^2}{4M^2}, \quad c_2 =\frac{c_\psi^4 v^2}{4M^2}~.
\end{equation}

In the Appendix we calculate the shifts in the electroweak precision
observables in terms of the parameters $c_1$ and $c_2$ defined above.
We perform a two parameter global fit to the precision electroweak
data given in Table~\ref{table} assuming a 115 GeV Higgs 
(as described in \cite{RSfit}) , and find
that $c_1$ and $c_2$ are tightly constrained as demonstrated in
Fig.~\ref{fit-fig}.  The best fit in $c_1$ and $c_2$ has 
$\chi^2\simeq 30.5$ with 23 observables.
(We note in passing that shifting the Higgs to 
heavier masses worsens the best $\chi^2$ fit to the data.)
We will translate these constraints into
bounds on $\tilde{g}$, $\tilde{g}'$ and the relevant mass scales,
after discussing the running of the gauge couplings.

\FIGURE[t]{
\epsfxsize=4.0in
\centerline{\epsfbox{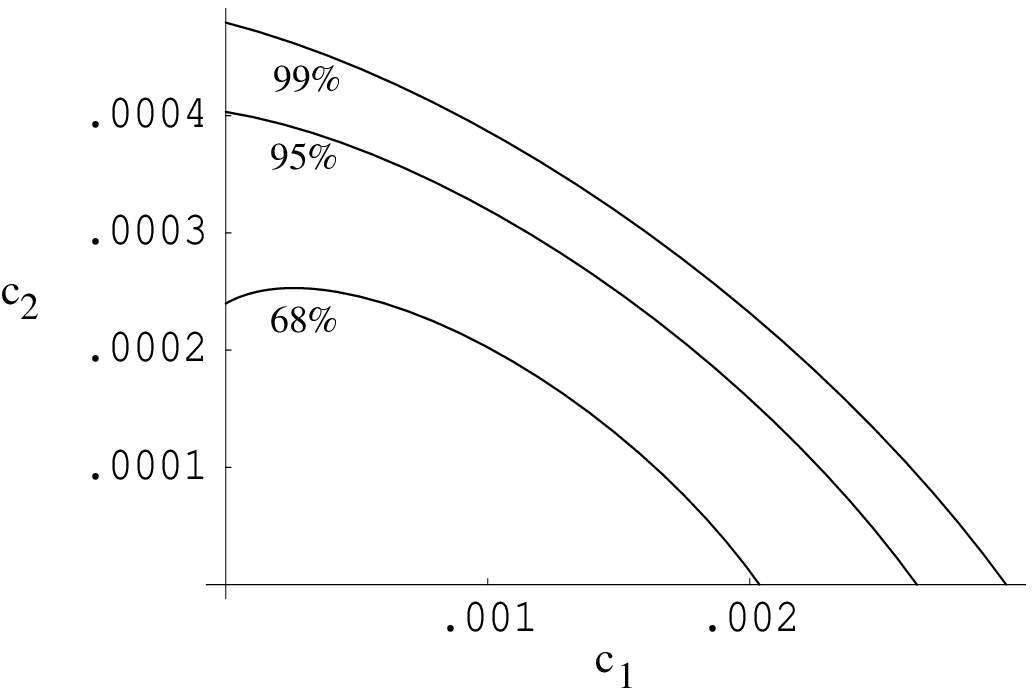}}
\caption{Confidence ellipses in $c_1$-$c_2$ space.}
\label{fit-fig}}

\section{Constraints on the High Energy Parameters}
\label{RG-sec}

We now relate the constraints on $c_1$ and $c_2$ to the parameters
of the SU(3)$_W$ $\times$ SU(2) $\times$ U(1) theory.  At the 
matching scale $M_u={\rm Max}[M_{W_H},M_{B_H}]$, 
the high energy gauge couplings are related
to the electroweak couplings through
\begin{eqnarray}
\frac{1}{g^2(M_u)} &=& \frac{1}{g_3^2} + \frac{1}{\tilde{g}^2} 
\label{match-g-eq} \\
\frac{1}{g^{\prime\,2}(M_u)} &=& \frac{3}{g_3^2} + \frac{1}{\tilde{g}^{\prime\,2}} 
\label{match-gp-eq} \; .
\end{eqnarray}
As long as $g_3\ll \tilde{g},\tilde{g}'$ one expects the prediction
$\sin^2 \theta_W=1/4$ to approximately be satisfied. This is  similar to
the approximate SU(5) unification in the models in~\cite{decgut}. 

There are four unknowns (three gauge couplings and the matching 
scale) with two constraint equations.  The electroweak couplings evaluated 
at the matching scale are related to the well-measured couplings at 
$M_Z$ through the one-loop renormalization group equations
\begin{equation}
\frac{1}{g_a^2(M_u)} = \frac{1}{g_a^2{(M_Z)}} - \frac{1}{8 \pi^2}
    \left[   b_a \ln \frac{M_u}{M_Z} + \tilde{b}_a \ln \frac{M_u}{m_t} 
           + (d_a + e_a) \ln \frac{M_u}{M_i} \right] \; ,
\end{equation}
where $b_a = (53/9,-22/6)$, $\tilde{b}_a = (17/18,1/2)$, 
$d_a = (-33, -11/3)$, and $e_a = (3/2,1/2)$ for 
$a=[{\rm U(1)}_Y,{\rm SU(2)}_L]$.  The last term represents the
threshold correction from two contributions.  The first ($d_a$) 
corresponds to the intermediate scale gauge bosons with mass 
$M_i = g_3 M/\sqrt{2}$ (that do not mix with the SM gauge bosons).
The second ($e_a$) corresponds to all of the components of the 
scalar field $\Sigma$ that are not eaten by $W_H$ or $B_H$.  This includes 
the four Goldstone bosons eaten by the intermediate scale gauge bosons
and four uneaten scalars that we assume have mass $M_i$.  We also note that
these sharp threshold corrections at the masses of heavy fields are only
approximate, but additional corrections do not significantly modify our
results.

The high energy theory is therefore completely determined by $\tilde{g}$ 
and $\tilde{g}'$.  Any given point in this two-parameter space has
a definite prediction for the matching scale $M_u$, the SU(3)$_W$
coupling, and $c_1$ and $c_2$.  It is straightforward
to determine the region of $\tilde{g}$-$\tilde{g}'$ space that is
excluded by large contributions to the electroweak precision observables,
which we show in Fig.~\ref{restrict-fig}.
\FIGURE[t]{
\centerline{
\epsfxsize=0.55\textwidth
\epsfbox{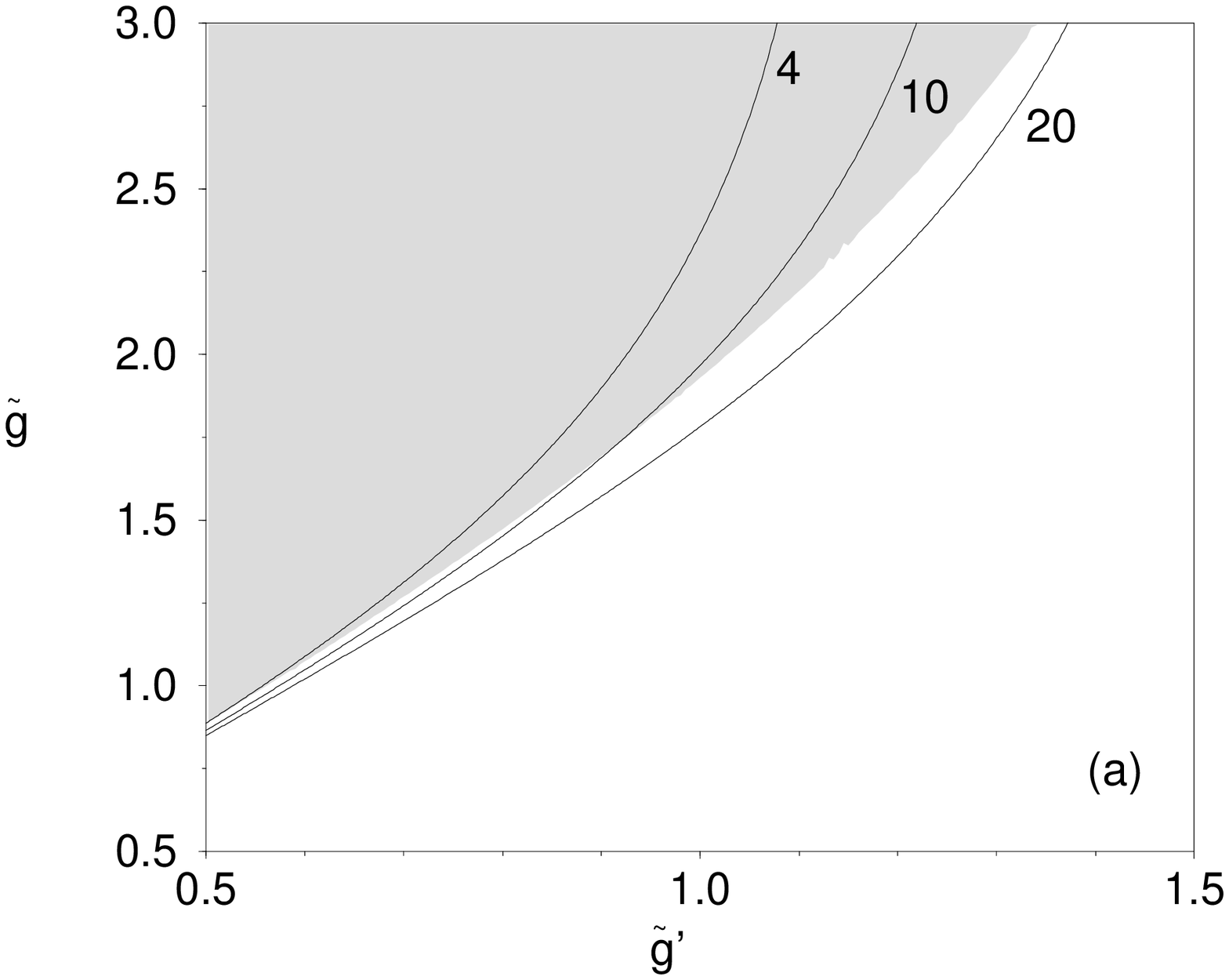}
\hfill
\epsfxsize=0.55\textwidth
\epsfbox{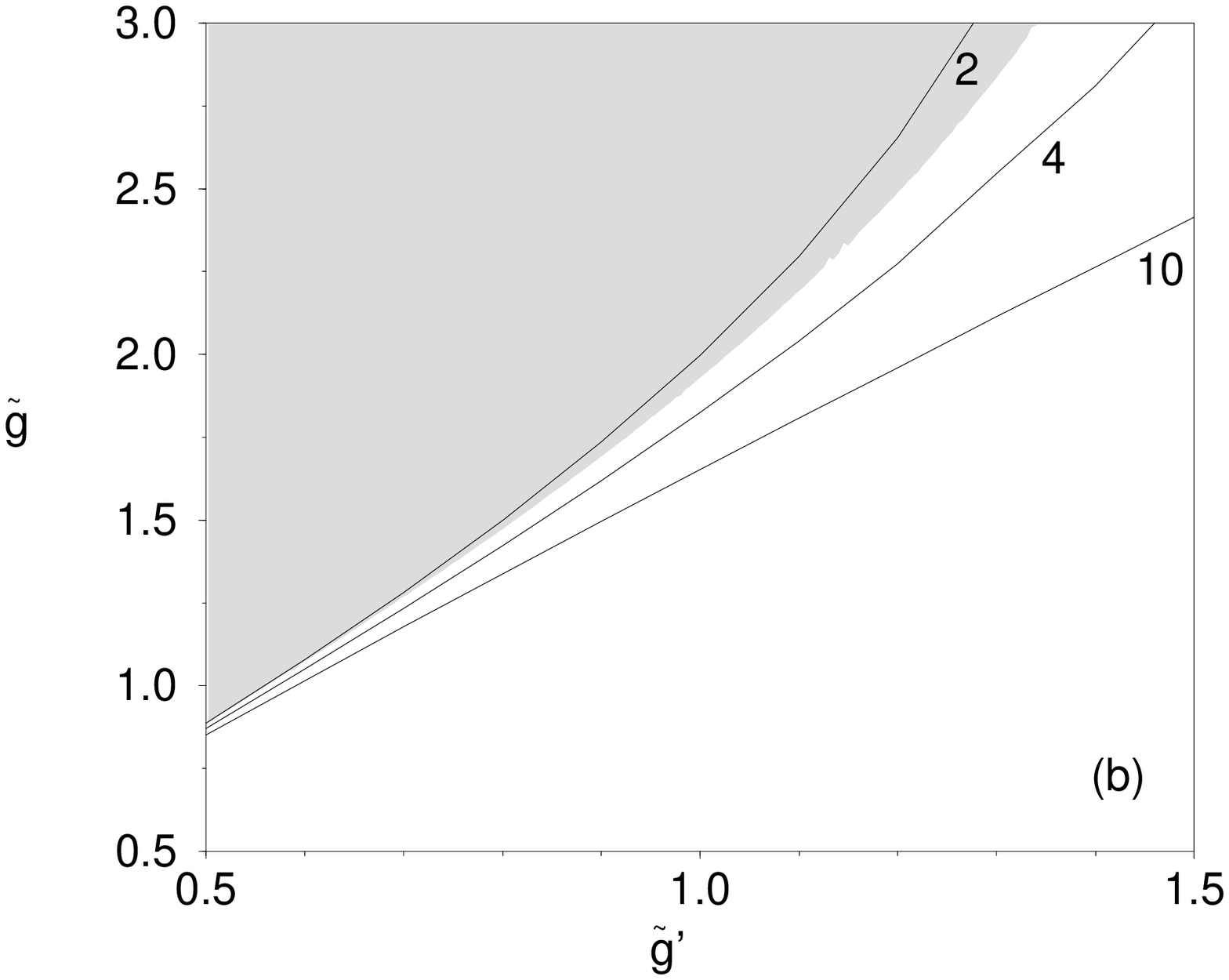}}
\caption{Region of $\tilde{g}$-$\tilde{g}'$ space that is excluded
(shaded region)
to 95\% C.L. by electroweak precision observables.  The solid lines 
correspond to (a) contours of the matching scale $M_u$ in TeV,
and (b) contours of the intermediate gauge boson masses $M_i$ in TeV.}
\label{restrict-fig}}
Several comments are in order.
Small couplings $\tilde{g}$, $\tilde{g}' \lsim 0.5$
are forbidden since there is no solution to the matching conditions
(\ref{match-g-eq})-(\ref{match-gp-eq}).  Couplings larger than $3$
are not shown since perturbation theory begins to break down.  
We find that the matching (or unification) scale must be larger than 
$4$ TeV throughout the physically acceptable region, and larger
than $11$ TeV for the region $\tilde{g},\tilde{g}' > 1$.

It is clear from Fig.~\ref{restrict-fig} that small changes in the 
high energy gauge couplings $\tilde{g}$, $\tilde{g}'$ result
in large changes to the matching scale.  Consider for example what 
is needed to obtain a matching scale of order $20$ TeV.  The region 
of interest comprises $0.5 \lsim \tilde{g}' \lsim 1.4$, within which
the \emph{a priori} independent coupling $\tilde{g}$ must be fine-tuned 
to approximately satisfy $\tilde{g} \simeq \sqrt{3} \tilde{g}'$.  
Along this line, the threshold corrections to $\sin^2 \theta_W(M_u)$ 
accidentally cancel out.  However, the
degree of fine-tuning needed for a fixed matching scale varies 
quite significantly with $\tilde{g}'$.  We can quantify this by 
calculating the fractional shift in the high energy couplings that
corresponds to a given fractional shift in the matching scale, shown 
in Fig.~\ref{finetune-fig}.
\FIGURE[t]{
\epsfxsize=0.60\textwidth
\centerline{\epsfbox{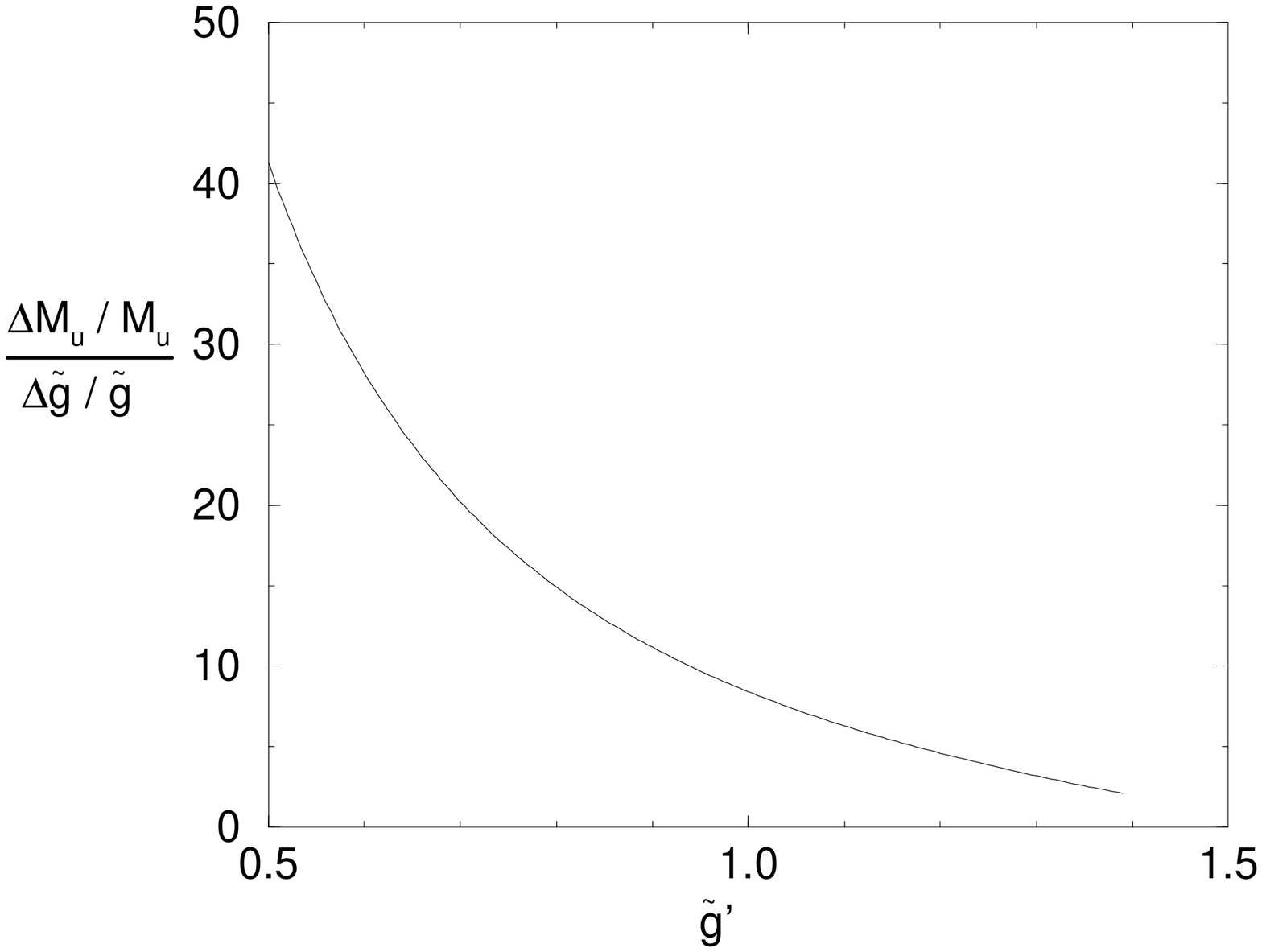}}
\caption{A measure of the fine-tuning of the high energy
gauge couplings.  The fractional change in the matching scale $M_u$
normalized to the fractional change in the SU(2) coupling $\tilde{g}$ 
is shown as a function of the U(1) coupling $\tilde{g}'$.  
For example, this means that holding $M_u$ fixed within 10\% requires 
tuning $\tilde{g}$ to 10\%/(y-axis value).  We used $M_u = 20$ TeV, 
although the result shown is quite insensitive to the value of the 
matching scale.}
\label{finetune-fig}}
For example, holding $M_u$ fixed within 10\% requires tuning $\tilde{g}$
to be within ($0.6\%$, $1.2\%$, $2.6\%$) for 
$\tilde{g}' =$ ($0.75$, $1$, $1.25$).  This means that arranging
that new physics be close to our bounds requires significant 
fine-tuning of the high energy gauge couplings.  This fine tuning can be 
relaxed, 
but only by simultaneously going to moderate gauge couplings 
$\tilde{g},\tilde{g}' > 1$ and matching scales well beyond our bounds, 
in the tens of TeV.

\section{Gauge Boson Phenomenology}
\label{pheno-sec}

Directly resolving the physics of the matching scale is clearly 
well out-of-reach for future colliders (Tevatron, LHC, LC).  
For the region of physical interest ($g,\,g'>1$) the heavy gauge bosons
which couple directly to quarks and leptons have masses 
$M_{W_H}>11$ TeV, $M_{B_H}>6$ TeV.
However, there are four gauge bosons, $A^{4}$-$A^{7}$ 
that are generally significantly lighter than the heaviest gauge bosons
$W_H$ and $B_H$.  This is shown in Fig.~\ref{restrict-fig}(b) where
the contours correspond to the intermediate gauge boson mass $M_i$.
$A^{4}$-$A^{7}$ do not couple to the SM fermions.
Hence, they do not contribute at tree-level to electroweak precision
observables, which is why they are permitted to be much lighter 
than the heavy mixed states $W_H$ and $B_H$.

The electroweak quantum numbers of $A^{4}$--$A^{7}$ are $(2,3/2)$ and 
$(2,-3/2)$, which means they have electric charges $\pm 2, \pm 1$.  
They can only
be produced in pairs through couplings to the standard model gauge bosons.
This is because they are SU(2)$_L$ doublets that
do not couple to the standard model matter fields, so that every  
interaction vertex must contain at least two of these fields to be
gauge invariant.  Therefore in order to observe these particles one would 
have to pair produce them.  For gauge couplings 
$\tilde{g},\tilde{g}' > 1$, their mass is larger than $2.5$ TeV, 
and so they are unlikely to be produced in sufficient quantities to
be detected above backgrounds at the LHC.  
Similarly, in the minimal model without 
any new matter charged under SU(3)$_W$, these gauge bosons cannot decay.  
Again this is a consequence of the coupling by pairs to the electroweak
gauge bosons that are triplets or singlets.  This ``doublet conservation''
is not broken by electroweak symmetry breaking since the Higgs does 
not couple to these gauge bosons.  The only other interaction
is potentially with the physical (3,0), (1,0) components of the 
$\Sigma$ scalar, 
but these also cannot be decay modes for the same reason.  Even adding
non-renormalizable operators to the theory would not change this situation.
The reason for this is conservation of the strange color $s$ (the third
color of the SU(3) group). This is not broken by the $\Sigma$ vev, and
gauge interactions leave the combination $s-\bar{s}$ invariant. This is
a result of discrete symmetries of the model
($A^\mu\rightarrow -A^\mu,\,
\Sigma\rightarrow -\Sigma$ and the SU(3) group element diag(-1,-1,1)), the
product of which remains unbroken.  For a decay
mode of the $A^{4,\ldots,7}$ gauge bosons $s-\bar{s}=\pm 1$ in the initial 
state, but is zero in any kinematically allowed final state, so there is
no allowed decay mode.

Electrically charged, stable particles can lead to
severe cosmological problems \cite{danger}. 
In fact, an order of magnitude estimate \cite{KT} shows
that these $A^{4,\ldots,7}$ gauge bosons would have a relic density 
much larger than the critical density, assuming their mass is of order 
$M_i=2$ TeV, 
and the annihilation cross section is of order $\alpha^2/M_i^2$. 
This is obtained using the simple estimate \cite{KT} for the relic density
\begin{equation}
\Omega h^2 \sim \frac{7.7 \times 10^{-38} \; {\rm cm}^2}{\langle 
\sigma v \rangle}~,
\end{equation}
where $\Omega$ is the fractional energy density compared to the 
critical density, $h$ is the Hubble constant in units of $100$ km/s/Mpc,
and $\langle \sigma v \rangle$ is the average annihilation cross section 
that we estimate to be of order $\alpha^2/M_i^2$. The result for 
$M_i = 2$ TeV is $\Omega \sim 15$.
Thus these particles may overclose the Universe.  But, there is an even
stronger experimental constraint on the relic density.  Stringent bounds 
on the relative abundances of new charged stable particles have been 
set by searches for heavy isotopes of ordinary nuclei.  The typical 
relative abundance obtained for these charged gauge bosons is of the 
order $n_{A^{4,\ldots,7}}/n_{nucleons} \sim 10^{-1}$.
Searches for heavy isotopes \cite{isotopesearch}, however, 
typically set a bound on this 
abundance of the order  $n_{A^{4,\ldots,7}}/n_{nucleons} < 
10^{-15} - 10^{-20}$.  This suggests that the presence of these stable 
charged gauge bosons $A^{4,\ldots,7}$ is excluded.

One could try to avoid this constraint by adding light matter charged 
under just the SU(3)$_W$, to which the charged gauge bosons could decay.  
The difficulty with this approach is that the 
new matter automatically has electroweak quantum numbers and is therefore 
not immune to experimental constraints.  One possibility is a vector-like
pair of SU(3)$_W$ triplets unrelated to the SM matter.  
But this is 
both theoretically unappealing (why is the vector-like triplet mass scale 
near the electroweak scale?) as well as being constrained by electroweak 
measurements. In addition, a stable particle would still remain
in the spectrum, since the conservation of the strange color introduced above 
requires
that the lightest of the particles charged under this symmetry be stable.
Generically, this particle will be charged under the unbroken U(1)$_{EM}$.
In this case the mass of the stable particle
could be much smaller than before ($\sim 100$ GeV instead
of TeV) and thus might not overclose the Universe. However the bound from 
isotope searches would be difficult to evade even in this case.

Another choice might be to assume 
some or all of the leptons are SU(3)$_W$ triplets.  This causes 
SU(2) $\times$ U(1) anomalies, and so yet more matter must be added 
to the model just to cure this problem.  
(Anomaly cancellation was also a generic problem in the
``ununified standard model'' \cite{Georgiununified}.)
Furthermore, new operators would exist in the low energy effective theory 
affecting the (SU(3)$_W$ triplet) lepton couplings once $A^{4}$-$A^{7}$ 
are integrated out.  There is every reason to expect electroweak
observables would then place as strong a constraint on the mass of these 
gauge bosons as we found on the mass of $W_H$ and $B_H$.  
This means that the lower bound on the matching scale would be increased
to of order 30 TeV.

Thus, we find that the ``minimal'' SU(3)$_W$ $\times$ SU(2) $\times$ U(1) 
idea has a potentially serious cosmological obstacle in the form of heavy, 
stable, charged gauge bosons.  The solutions to
this problem involve adding new light matter that is charged under 
the SU(3)$_W$ group, but this new sector is expected to be strongly 
constrained by electroweak precision measurements.

\section{Conclusions}
\label{conclusions-sec}

We have studied constraints on the recently proposed
SU(3)$_W\times$SU(2)$\times$U(1) electroweak model.  By integrating out the
heavy gauge bosons we calculated the low energy effective action in
this model.  We identified the corrections to the masses and couplings of
standard model fields.  This 
allowed us to calculate the corrections to the electroweak observables in
this model, from which we performed a global fit to current experimental
data.  By fixing the value of the electric charge and $\sin^2\theta_W$,
the model was specified in terms of just two parameters.  We found the excluded
region in the space of SU(2) and U(1) couplings, and found that for the
physically interesting region the unification scale is bounded to be larger
than 11 TeV.

We also pointed out there are stable multi-TeV scale gauge bosons
that are electrically charged, leading to cosmological difficulties.  
These gauge bosons would have been produced
in the early universe, and we found that their present-day relic density 
would be 
larger than the critical density, and so would overclose the Universe.  
Furthermore, there are considerably stronger experimental constraints 
on the relic density of stable, charged particles coming from
searches for heavy isotopes of nuclei.  This suggests that 
modifications to the minimal model are needed to allow the gauge boson
to decay.  The simplest idea of allowing leptons or additional
matter to transform under SU(3)$_W$ lead to further model-building
difficulties ({\em e.g.} anomaly cancellation) and experimental constraints
(new contributions to precision electroweak observables).

\section*{Acknowledgments}
We thank J.~Erler for providing us the SM predictions used in our fit.
GDK thanks T.~Plehn and D.~Zeppenfeld for discussions.  JE and JT thank
A.~de~Gouvea for comments.
The research of C.C. is supported in part by the NSF, and in part by the
DOE OJI grant DE-FG02-01ER41206. The research of J.E. and J.T. is supported
by the US Department of Energy under contract W-7405-ENG-36.
The research of G.D.K. is supported by the US Department of Energy
under contract DE-FG02-95ER40896.

\section*{Appendix A:  Predictions for  Electroweak Observables}
\renewcommand{\theequation}{A.\arabic{equation}}

In this appendix we give the predictions for the shifts in 
the electroweak precision observables as a result of the
new gauge dynamics SU(3)$_W$ $\times$ SU(2) $\times$ U(1).
The electroweak observables depend on only two parameters,
$c_1$ and $c_2$.  Using the results given in~\cite{Sformulas,Burgess} 
as well as the low-energy $\nu e$ couplings:
\begin{equation}
 g_{eV}(\nu e \rightarrow \nu e) = 2 \left(s^2 - \frac{1}{4}\right)~, 
\qquad
 g_{eA}(\nu e \rightarrow \nu e) = -\frac{1}{2}~. \nonumber
\end{equation}
we find the following results:
\begin{eqnarray}
\Gamma_Z &=& \left( \Gamma_Z \right)_{SM} \left(1 -0.89 c_1+0.17 c_2\right) 
 \nonumber \\
R_e &=& \left( R_e \right)_{SM} \left(1 +0.082 c_1 +0.91 c_2\right) 
 \nonumber \\
R_\mu &=& \left( R_\mu \right)_{SM} \left(1   +0.082 c_1 +0.91 c_2\right)
 \nonumber \\
R_\tau &=& \left( R_\tau \right)_{SM} \left(1 +0.082 c_1 +0.91 c_2\right)
 \nonumber \\
\sigma_h &=& \left( \sigma_h \right)_{SM} \left(1 -0.0087 c_1-0.096 c_2 \right)
 \nonumber \\
R_b &=& \left( R_b \right)_{SM} \left(1 -0.018 c_1-0.20 c_2 \right) 
 \nonumber \\ 
R_c &=& \left( R_c \right)_{SM} \left(1 +0.035 c_1 +0.39 c_2\right) 
 \nonumber \\
A_{FB}^e &=& \left( A_{FB}^e \right)_{SM}  +0.18 c_1 +2.0 c_2\nonumber \\
A_{FB}^\mu &=& \left( A_{FB}^\mu \right)_{SM}  +0.18 c_1 +2.0 c_2\nonumber \\
A_{FB}^\tau &=& \left( A_{FB}^\tau \right)_{SM}+0.18 c_1 +2.0 c_2\nonumber \\
A_{\tau}(P_\tau) &=& \left( A_{\tau}(P_\tau) \right)_{SM} +0.78 c_1 +8.6 c_2 
\nonumber \\
A_{e}(P_\tau) &=& \left( A_{e}(P_\tau) \right)_{SM} +0.78 c_1 +8.6 c_2
\nonumber \\
A_{FB}^b &=& \left( A_{FB}^b \right)_{SM} +0.54 c_1 +6.0 c_2 \nonumber \\
A_{FB}^c &=& \left( A_{FB}^c \right)_{SM} +0.42 c_1 +4.7 c_2 \nonumber 
\end{eqnarray}
\begin{eqnarray}
A_{LR} &=& \left( A_{LR} \right)_{SM} +0.78 c_1 +8.6 c_2 \nonumber \\
M_W &=& \left( M_W \right)_{SM} \left(1+0.43 c_1+1.4 c_2 \right) \nonumber \\
M_W/M_Z &=& \left( M_W/M_Z \right)_{SM} \left(1+0.43 c_1+1.4 c_2 \right) 
 \nonumber \\
g_L^2(\nu N \rightarrow \nu X) &=& 
\left( g_L^2(\nu N \rightarrow \nu X) \right)_{SM} +0.25 (c_1+c_2) \nonumber \\
g_R^2(\nu N \rightarrow \nu X) &=& 
\left( g_R^2(\nu N \rightarrow \nu X) \right)_{SM} -0.085 (c_1+c_2)\nonumber \\
g_{eV}(\nu e \rightarrow \nu e) &=& 
\left( g_{eV}(\nu e \rightarrow \nu e) \right)_{SM} -0.66 (c_1+c_2)\nonumber \\
g_{eA}(\nu e \rightarrow \nu e) &=& 
\left( g_{eA}(\nu e \rightarrow \nu e) \right)_{SM} \nonumber \\
Q_W(Cs) &=& \left( Q_W(Cs) \right)_{SM}+ 73 (c_1+c_2) \nonumber
\end{eqnarray}
We also give in Table~\ref{table} 
the experimental data \cite{LEPEWG,ErlerLang}
and the SM predictions used for our fit.
\begin{table}[hb]
\begin{center}
\begin{tabular}{|c|c|c|}\hline
Quantity & Experiment & SM($m_h=115$ GeV) \\ \hline 
$\Gamma_Z$ & 2.4952 $\pm$ 0.0023 & 2.4965 \\
$R_e$ & 20.8040 $\pm$ 0.0500 & 20.7440 \\
$R_\mu$ & 20.7850 $\pm$ 0.0330 & 20.7440 \\ 
$R_\tau$ & 20.7640 $\pm$ 0.0450 & 20.7440 \\ 
$\sigma_h$ & 41.5410 $\pm$ 0.0370 & 41.4800 \\ 
$R_b$ & 0.2165 $\pm$ 0.00065 & 0.2157  \\
$R_c$ & 0.1719 $\pm$ 0.0031 & 0.1723 \\
$A_{FB}^e$ & 0.0145 $\pm$ 0.0025 & 0.0163 \\
$A_{FB}^\mu$ & 0.0169 $\pm$ 0.0013 & 0.0163 \\
$A_{FB}^\tau$ & 0.0188 $\pm$ 0.0017 & 0.0163 \\
$A_{\tau}(P_\tau)$ & 0.1439 $\pm$ 0.0041 & 0.1475 \\ 
$A_{e}(P_\tau)$ & 0.15138 $\pm$ 0.0022 & 0.1475 \\
$A_{FB}^b$ & 0.0990 $\pm$ 0.0017 & 0.1034 \\
$A_{FB}^c$ & 0.0685 $\pm$ 0.0034 & 0.0739 \\
$A_{LR}$ & 0.1513 $\pm$ 0.0021 & 0.1475 \\
$M_W$ & 80.450 $\pm$ 0.039 & 80.3890 \\
$M_W/M_Z$ & 0.8822 $\pm$ 0.0006 & 0.8816 \\
$g_L^2(\nu N \rightarrow \nu X)$ & 0.3020 $\pm$ 0.0019 & 0.3039 \\
$g_R^2(\nu N \rightarrow \nu X)$ & 0.0315 $\pm$ 0.0016 & 0.0301 \\
$g_{eA}(\nu e \rightarrow \nu e)$ & -0.5070 $\pm$ 0.014 & -0.5065 \\
$g_{eV}(\nu e \rightarrow \nu e)$ & -0.040 $\pm$ 0.015 & -0.0397 \\
$Q_W(Cs)$ & -72.65 $\pm$ 0.44 & -73.11 \\ 
$m_{\rm top}$ & $174.3 \pm 5.1$ & $176.3$ \\ \hline
\end{tabular}
\end{center}
\caption{The experimental results~\cite{ErlerLang,LEPEWG}
and the SM predictions for the various
electroweak precision observables used for the fit. The SM predictions 
are for $m_h=115$ GeV and $\alpha_s=0.120$ and  
calculated~\cite{Erler} using GAPP~\cite{GAPP}.}
\label{table}
\end{table}

\end{document}